\documentclass[preprint,prl,amsmath]{revtex4}
\usepackage{graphicx}
\begin{document}
\draft
  
\title{Control of black hole evaporation?}
  
\author{Doyeol Ahn\thanks{Also with Department of Electrical
    Engineering, University of Seoul, Seoul 130-743,
    Korea}\thanks{E-mail; dahn@uoscc.uos.ac.kr}}
  
\address{Institute of Quantum Information Processing and Systems,
  University of Seoul, \\
  90 Jeonnong, Tongdaemonn-ku, Seoul, Korea\\
E-mail: dahn@uos.ac.kr; davidahn@hitel.net\\
 }

\date{\today}
  
\begin{abstract}
Contradiction between Hawking's semi-classical arguments and the string theory on the evaporation of a black hole
has been one of the most intriguing problems in fundamental physics. 
A final-state boundary condition inside the black hole was proposed by Horowitz and Maldacena to resolve this contradiction. 
We point out that the original Hawking effect can also be regarded as a separate boundary condition at the event horizon for this scenario. 
Here, we found that the change of the Hawking boundary condition may affect the information transfer from the initial collapsing matter
to the outgoing Hawking radiation during the evaporation process and as a result the evaporation process itself, significantly.  
\end{abstract}

\pacs{PACS numbers: 03.67.-a, 89.70.+c}

\maketitle

\pagebreak

The Hawking effect \cite{hawk1,hawk2} on the information loss in a black hole has been a serious challenge to modern physics because it contradicts the basic principles of quantum mechanics. Hawking's semi-classical argument predicts that a process of black hole formation and evaporation is not unitary \cite{hawk3}. On the other hand, there is some evidence in string theory that the formation and the evaporation of black hole is a unitary process \cite{HM}. Nonetheless, the Hawking effect, discovered nearly 30 years ago, is generally accepted very credible and considered as would be an essential ingredient of the yet unknown correct theory of quantum gravity. 

Previously, Horowitz and Maldacena (HM) proposed a final-state boundary condition \cite{HM} to reconcile the unitarity of the black hole evaporation with Hawking's semi-classical reasoning.  The essence of HM proposal is to impose a unique final boundary condition at the black hole singularity such that no information is absorbed by the singularity. The final boundary state is a maximally entangled state of the collapsing matter and the infalling Hawking radiation. The projection of final boundary state at the black hole singularity collapses the state into one associated with the collapsing matter and transfer the information to the outgoing Hawking radiation. The HM model is further refined, by including the unitary interactions between the collapsing matter and the infalling Hawking radiation \cite{Preskill}, with a random purification of the final boundary state \cite{Lloyd1}. One of the critical assumptions in the 
HM proposal is that the internal quantum state of the black hole can be represented by a maximally entangled state of the collapsing matter 
and the infalling Hawking radiation. This ansatz is important because the final state boundary condition of the HM proposal is based on this maximally 
entangled internal quantum state \cite{HM, Preskill}. Recently, the author proved the HM ansatz for the special case of a collapsing gravitational shell 
inside the Schwarzschild black hole \cite{Ahn}. 

In the HM model, the boundary state outside the event horizon is assumed to be the Unruh vacuum state \cite{Unruh, Wald}. 
As a matter of fact, Hawking's original discovery can be regarded as imposing a boundary condition at the event horizon. 
The author would like to denote it as the Hawking boundary condition (HBC) in contrast with the final-state boundary condition (FBC) proposed by HM (Fig. 1). 
The HBC dictates that the quantum states inside and outside the event horizon of the black hole are maximally entangled. 
\pagebreak
\begin{figure}[htbp]

 \centering
 \includegraphics[width=0.5\textwidth,angle=-90]{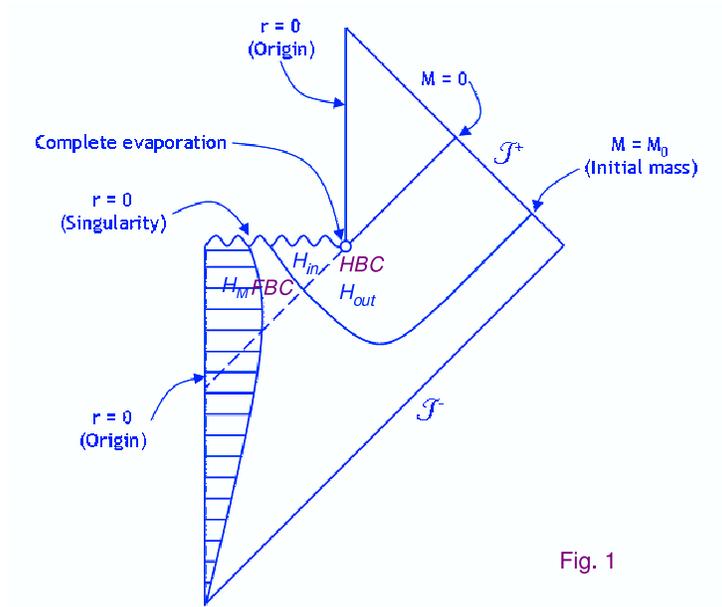}
  \caption{Penrose diagram of for the black hole formation and evaporation processes \cite{Wald}. The HBC denotes the Hawking boundary conditions at the event horizon and the FBC denotes the final-state boundary condition inside the black hole.  $J^{+}$ and $J^{-}$  are the future and the past null infinity, respectively.}
\end{figure}
Significance of this scenario is that the black hole formation and the evaporation process can be put into a unified picture by combining the HBC together with the FBC.

It would be an interesting question to ask whether the black hole evaporation process 
will be affected by the boundary condition at the event horizon. 
The boundary condition on the event horizon would affect the final state projection 
because the quantum states inside and outside the event horizon are entangled by the HBC. 
The purpose of this paper is to demonstrate that the final state boundary condition (FBC) of Horowitz and Maldacena 
necessarily implies that the evaporation process  depends on the boundary condition on the horizon. 
The author proceeds by assuming that (a) the original vacuum outside of a black hole evolves into 
a maximally entangled state on $H_{in}$  and  $H_{out}$ 
and then (b) the interior state of the black hole is also a maximally entangled state on $H_{in}$  and $H_{M}$ , 
where $H_{M}$  is the Hilbert space for the collapsing matter.  
The FBC is then applied to the latter state, and the outgoing radiation is obtained by projection of this onto the former state. 
However if one took a state in which a single boson is excited outside of the black hole as the Hawking boundary condition 
then it too could be written as a maximally entangled state on $H_{in}$  and $H_{out}$ 
 - but the state of outgoing radiation that one would obtain via the FBC is different from the state obtained from the vacuum state.  
Hence one obtains the result that the final outgoing particle state for a black hole evaporation is dependent on the Hawking boundary condition. 

We assume that the quantum state of the collapsing matter belongs to a Hilbert space $H_{M}$ with dimension $N$  and $|n\rangle_{M}~$ be the initial quantum state of the collapsing matter. 
It is also assumed that  $|n\rangle_{M}$ belongs to the set of orthonormal basis $\{|{l}\rangle_{M}\}$  for $H_{M}$. 
The Hilbert space of fluctuations on the background spacetime for black hole formation and evaporation 
is separated into $H_{in}$  and $H_{out}$  which contain quantum states localized inside and outside the event horizon, respectively. 
In HM proposal, the HBC is assumed to be the Unruh vacuum state $|\Phi_0\rangle_{in\otimes out}$   belonging to $H_{in}\otimes H_{out}~$  
in a micro-canonical form \cite{HM, Preskill, Lloyd1, Ahn}:

\begin{equation}
|\Phi_0 \rangle_{in\otimes out}={1\over\sqrt{N}}\sum_{l=1}^N|{l}\rangle_{in}\otimes|{l}\rangle_{out}~,
\end{equation}
where $\{|{l}\rangle_{in}\}$  and  $\{|{l}\rangle_{out}\}$   are orthonormal bases for  $H_{in}$ and $H_{out}$ , respectively. 
The final-state boundary condition (FBC) imposed at the singularity requires 
a maximally entangled quantum state in  $H_{M}\otimes H_{in}$ which is called the final boundary state and is given by

\begin{equation}
{}_{M\otimes in}\langle \Psi|={1\over \sqrt{N}}\sum_{l=1}^N {}_M\langle {l}|\otimes {}_{in}\langle {l} |(S \otimes I)~,
\end{equation}
where S is a unitary transformation.  
The initial matter state $|n\rangle_{M}$ evolves into a state in $H_{M}\otimes H_{in} \otimes H_{out}~$  under HBC, 
which is given by $|\Psi_0 \rangle_{M \otimes {in} \otimes {out}}=|n\rangle_{M} \otimes |\Phi_0 \rangle_{{in} \otimes {out}}~$. 
Then the transformation from the quantum state of collapsing matter to the state of outgoing Hawking radiation is given by the following final state projection \cite{Lloyd1}

\begin{equation}
|\phi_0\rangle_{out}={}_{M \otimes {in}} \langle \Psi | \Psi_0 \rangle_{M \otimes {in} \otimes {out}}=\sum_{i} {}_{M} \langle i|S|n \rangle_{M} |i \rangle_{out}~,
\end{equation}
where right side of Eq. (3) is properly normalized. Let's assume that the orthonormal bases $\{ |i\rangle_{out} \}$  and   $\{ |{l}\rangle_{M} \}$ are related by the unitary transformation $T'$, then one can easily show that  ${}_{out} \langle i|T|{n} \rangle_{M}=\delta_{i,n}~$. 
The quantum state of the collapsing matter is transferred to the state of the outgoing Hawking radiation with the fidelity defined by

\begin{equation}
{f}_0=|{}_{out} \langle \phi_0 |T'| {n} \rangle_{M}|^2=|{}_{M} \langle {n} |S| {n} \rangle_{M}|^2.
\end{equation}

I would like to note that we can also regard  $T'$ as a tunnelling Hamiltonian \cite{Bardeen} 
and the evaporation rate will be proportional to  ${2\pi \over \hbar }{f}_0$.

\begin{figure}[htbp]

 \centering
 \includegraphics[width=0.65\textwidth,angle=-90]{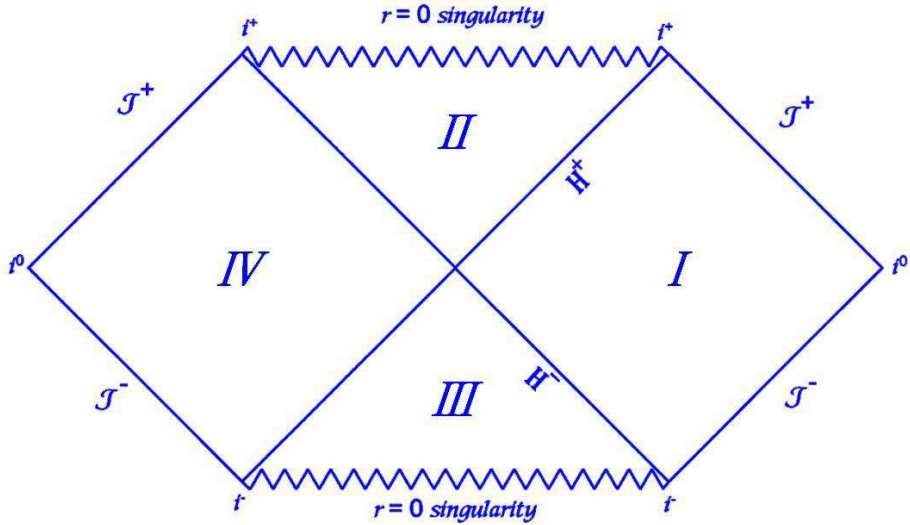}
  \caption{The Kruskal extension of the Schwarzschild spacetime \cite{Wald, Davies}. 
In the region $I$, null asymptotes $H_{+}$ and $H_{-}$ act as the future and the past horizon, respectively. The boundary lines labeled  $J^{+}$ and $J^{-}$  are the future and the past null infinity, respectively, and $i^{0}$ is the spacelike infinity.}
\end{figure}

Now we consider the case of imposing the Unruh excited state as the HBC on the black hole evaporation problem.
The Unruh vacuum state is evolved unitarity from the vacuum state defined in the far past such as the past null infinity $J^{-}$ \cite{HM} (Fig. 2). Let's denote $| 0_{-}\rangle$ and $a$ as the vacuum state and the annihilation operator of a particle, respectively, at the past null infinity \cite{Parker}, and $V$ is the unitary operator responsible for the evolution of the vacuum state 
$| 0_{-}\rangle~$.
Then, we have
\begin{equation}
|\Phi_0\rangle_{in\otimes out}=V| 0_{-}\rangle
\end{equation}
with $a| 0_{-}\rangle=0$.

We now define the one-particle state with a finite energy at the past null infinity by $|1_{-} \rangle$ which is given by
\begin{equation}
|1_{-} \rangle=a^{\dagger} |0_{-} \rangle.
\end{equation}
Then the unitary evolution of this one-particle state at the future null infinity is given by
\begin{eqnarray}
V|1_{-} \rangle & = & Va^{\dagger} |0_{-} \rangle \nonumber \\
& = & Va^{\dagger} V^{-1} V|0_{-} \rangle \nonumber \\
& = & Va^{\dagger} V^{-1}|\Phi_0\rangle_{in\otimes out}.
\end{eqnarray}
The Unruh excited state at the future null infinity is then obtained by applying the Bogoliubov transformation \cite{Wald, Fuentes, Alsing, Davies} on the Unruh vacuum state and is given by

\begin{eqnarray}
|\Phi_1\rangle_{in\otimes out} & = & V|1_{-} \rangle \nonumber \\
& = & C(N)(\sqrt{N}b^{\dagger}_{out}-\sqrt{N-1}b_{in})|\Phi_0\rangle_{in\otimes out} \nonumber \\
& = &{\sqrt{2 \over {N^2-1}}}\sum_{i=1}^N \sqrt{i+1}|i\rangle_{in}\otimes|i+1\rangle_{out} \nonumber \\
& = & {\sqrt{2 \over{N^2-1}}}b^{\dagger}_{out} \sum_{i} |i \rangle_{in} \otimes |i \rangle_{out} \nonumber \\
& = & {\sqrt{2 \over{N^2-1}}}b^{\dagger}_{out}|\Phi_0 \rangle_{{in} \otimes {out}} ~,
\end{eqnarray}
where $b^{\dagger}_{out}$ is the boson creation operator outside the black hole, $b_{in}$ is the boson annihilation operator inside the black hole, and $C(N)$ is the normalization factor. At later times one can choose a spacelike slice that goes through the future horizon. The horizon divides this spacelike hypersurface into two parts, one inside and one outside the horizon \cite{HM}. As a result, the Hilbert space can be separated into two parts, $H_{in}~$ and $H_{out}$ which contain wave functions localized inside or ouside the horizon. Here the operator $b^{\dagger}_{out}~$ corresponds to the operater $b^{\dagger}~$ of Parker's work \cite{Parker} which operates on a particle at the future null infinity. On the other hand,
the operator $b_{in}~$ annihilates a particle incoming at the future horizon $H^{+}~$ corresponding to the operator $c$ of reference \cite{Parker}.
In Eq. (8), we assumed the density operator  $ \rho_{1 \ out} \equiv Tr_{in} (|\Phi_1 \rangle_{in \otimes out} \langle \Phi_1| )~$ has a unit trace.

Here the mathematical form of the Bogoliubov transformation 
in a micro-canonical ensemble $Va^{\dagger} V^{-1}=C(N)(\sqrt{N}b^{\dagger}_{out}-\sqrt{N-1}b_{in})~$ is taken after the Bogoliubov transformation, $a^{\dagger}_K=cosh(r_{\omega})b^{\dagger}_{out}-sinh(r_{\omega})b_{in}~$ 
for the Schwarzschild-Kruskal spacetime \cite{Ahn, Wald3, Parker}, where the subscript $K$ denotes the Kruskal spacetime,
$\omega$ the postive frequency of the normal mode, and $r_{\omega}$ denotes the squeezing parameter \cite{Barnett}. In the latter, the Bogoliubov transformation was derived from an explicit wave packet basis and an analysis was carried out by taking into account the real physical properties of the state. 
One may regard the Bogoliubov transformation for the micro-canonical ensemble used in the paper as a limiting case of the Bogoliubov transformation in the Schwarzschild-Kruskal spacetime for which a timelike Killing vector associated with Kruskal coordinates is well defined for certain regions or 
boundary lines. But they are not unique. For example, $\frac{\partial}{\partial \overline{u}}~$ is a timelike Killing vector on $H_{-}~$ \cite{Davies}. In our simplified model, the momenta or the frequencies of the particles are not specified explicitly but we assume that the Unruh excited state contains a particle of finite energy. 
From the above equation, one can see that $|\Phi_1 \rangle_{in \otimes out}~$  is effectively a single boson excited state, i.e., a state 
in which a particle is created outside the black hole.  
Then the initial matter state $|n \rangle_{M} $ evolves into a state $|\Psi_1 \rangle_{M \otimes in \otimes out}$ in $H_{M} \otimes H_{in} \otimes H_{out}~$  under the HBC, which is given by  $|\Psi_1 \rangle_{M \otimes in \otimes out}=|n \rangle_{M} \otimes |\Phi_1 \rangle_{in \otimes}$. 
The final state projection yields

\begin{eqnarray}
|\phi_1\rangle_{out} & = &{}_{{M} \otimes {in}}\langle \Psi|\Psi_1 \rangle_{M \otimes {in} \otimes {out}} \nonumber \\
& = &{1\over \sqrt{\sum_{i}(i+1)|{}_{M}\langle i|S|n \rangle_{M}|^2}}\sum_{i}\sqrt{i+1} {}_M \langle i|S|n \rangle_{M}|i+1 \rangle_{out}.
\end{eqnarray}

The fidelity of information transfer $f_1$  from the collapsing matter to the out-going Hawking radiation is given by

\begin{eqnarray}
f_1 & = & |{}_{out}\langle \phi_1|T'|n \rangle_{M}|^2 \nonumber \\
& = & \frac {n|{}_{M}\langle n-1|S|n \rangle_{M}|^2} {\sum_{i}(i+1)|{}_{M}\langle i|S|n \rangle_{M}|^2}.
\end{eqnarray}

Gottesman and Preskill considered the generalization of HM proposal by considering the unitary interaction $U$ acting on $H_{M}\otimes H_{in}$ , i.e., the Hilbert space of the collapsing matter and the infalling Hawking radiation [5].  In their generalization, the final boundary state is given by

\begin{equation}
{}_{M\otimes in}\langle \Phi|={1\over \sqrt{N}}\sum_{l=1}^N {}_M\langle {l}|\otimes {}_{in}\langle {l} |(S \otimes I)U~,
\end{equation}
where the unitary transformation $U$ takes into account the interactions of the collapsing matter with the quantum field fluctuations after the horizon crossing but before the arrival at the singularity. Then one might ask the following question: Is the Unruh excited state given by eq. (8) essentially a special case of what one would get upon the unitary interaction between the collapsing matter and the Hawking radiation given by the generalized HM model given by eq. (11)? Mathematically, $U$ is acting on $H_{M}\otimes H_{in}$  and the HBC is prescribed on $H_{in}\otimes H_{out}$  so they are operating on different Hilbert spaces. Moreover, the HBC is prepared in the infinite past and logically U is supposed to be turned on after the HBC is prescribed. In order to answer this, we consider the following transformations;

\begin{eqnarray}
T_1 & = & {}_{M \otimes in} \langle \Psi |\Phi_1 \rangle_{in \otimes out} \nonumber \\
& = & \sqrt{\frac{2}{N(N^2-1)}}\sum_{l}\sqrt{{l}+1}|{l}+1\rangle_{out}{}_{M}\langle{l}|S
,
\end{eqnarray}

and

\begin{eqnarray}
T_2 & = & {}_{M \otimes in} \langle \Phi |\Phi_0 \rangle_{in \otimes out} \nonumber \\
& = & \frac{1}{N}\sum_{i,l}|i\rangle_{out}{}_{M}\langle{l}|S({}_{in}\langle{l}|U|i\rangle_{in}).
\end{eqnarray}
In eqs. (12) and (13), two states belong to $H_{in}$  are contracted with each other. In order to make $T_1$  and $T_2$  equivalent to each other, 
the matrix $U$ should be of the following form;

\begin{equation}
U=I_{M} \otimes U_{in}
\end{equation}

where

\begin{equation}
(U_{in})_{li}=\sqrt{\frac{2N({l}+1)}{N^2-1}}\delta_{{l}+1,i}
\end{equation}

Now the question is whether $U$ is unitary or more specifically, $U_{in}$  is unitary. 
From, $(U^{\dagger}_{in}U_{in})_{mn}=n\sqrt{\frac{2N}{N^2-1}}\delta_{m,n}~$ , it is obvious both $U_{in}$  and $U$ are not unitary. 
The Unruh vacuum state $|\Phi_0\rangle_{in\otimes out}$  and the excited state $|\Phi_1\rangle_{in\otimes out}$  
which comprise the HBCs  in $H_{in} \otimes H_{out}$  are not unitarily equivalent to each other and 
therefore the unitary transformation $U$ which makes $T_1$  and  $T_2$ equivalent does not exist.
Exact calculations of  $f_0$ and $f_1$  require a detailed knowledge of the unitary transformation $S$, 
so the direct comparison would be difficult at present. 
However, we can make a rough estimation using a refined HM model, which employs a random pure state as the FBC [6]

\begin{equation}
|\Psi\rangle_{M\otimes in}=\sum_{l}\lambda_{l}|{l}\rangle_{M}\otimes|{l}\rangle_{in}~,
\end{equation}
where $\lambda_{l}$  is the Schmidt coefficient for random state whose distribution is presumed to be known \cite{Lloyd2}.  
Random FBC also takes into account the stochastic interaction of the collapsing matter and the infalling Hawking radiation \cite{Preskill}. 
Substituting Eq. (16) into Eq. (3) yields, 

\begin{eqnarray}
{}_{M\otimes in}\langle \Psi|\Psi_0 \rangle_{M \otimes in \otimes out} & = & \sum_{l,j}\lambda^{\ast}_{l} {}_M \langle{l}| \otimes {}_{in}\langle{l}|n\rangle_{M}\frac{1}{N}|j\rangle_{in} \otimes |j\rangle_{out} \nonumber \\
& = & \frac{\lambda^{\ast}_{n}}{\sqrt{N}}|n\rangle_{out} \nonumber \\
& = & |\tilde{\phi_0}\rangle_{out}.
\end{eqnarray}

By normalizing $|\tilde{\phi_0}\rangle_{out}$ , 
we obtain the state of outgoing Hawking radiation, which is given by
\begin{equation}
|\phi_0\rangle_{out}=\frac{\lambda^{\ast}_{n}}{\sqrt{|\lambda_n|^2}}|n\rangle_{out}.
\end{equation}
Then the fidelity of information transfer for the Unruh vacuum state is given by

\begin{equation}
f_0=|{}_{out}\langle\phi_0|T'|n\rangle_{M}|^2=\frac{|\lambda_n|^2}{|\lambda_n|^2}=1.
\end{equation}
Likewise, by substituting eq. (16) into eq. (8), we obtain
\begin{eqnarray}
{}_{M\otimes in}\langle \Psi|\Psi_1 \rangle_{M \otimes in \otimes out} & = & \sum_{l,j}\lambda^{\ast}_{l} {}_M \langle{l}| \otimes {}_{in}\langle{l}|n\rangle_{M}\sqrt{\frac{2(J+1)}{N^2-1}}|j\rangle_{in} \otimes |j+1\rangle_{out} \nonumber \\
& = & \lambda^{\ast}_{n}\sqrt{\frac{2(j+1)}{N^2-1}}|n+1\rangle_{out} \nonumber \\
& = & |\tilde{\phi_1}\rangle_{out}.
\end{eqnarray}
By normalizing $|\tilde{\phi_1}\rangle_{out}$, we obtain the state of outgoing Hawking radiation, which is given by
\begin{equation}
|\phi_1\rangle_{out}=\frac{\lambda^{\ast}_{n}}{\sqrt{|\lambda_n|^2}}|n+1\rangle_{out},
\end{equation}
which is orthogonal to $|\phi_0\rangle_{out}$.

The fidelity of information transfer for the Unruh excited state is then given by 
\begin{equation}
f_1=|{}_{out}\langle\phi_1|T'|n\rangle_{M}|^2=(\frac{\lambda^{\ast}_{n}}{|\lambda_{n}|})^2|{}_{out}\langle n+1|T'|n\rangle_{M}|^2=0.
\end{equation} 
In Eq. (19) and Eq. (22), we assumed $|\Phi_0\rangle_{in \otimes out}$  and $|\Phi_1\rangle_{in \otimes out}$  respectively as the HBC. 
In this model, the fidelity for the information transfer from the collapsing matter to the outgoing Hawking radiation is zero. 
As a result the evaporation rate may be also suppressed when the Unruh excited state is taken as the HBC.
 
In general, the fidelity  $f_1$ would be in the range 
\begin{equation}
0\leq f_1 \leq \frac{n|{}_{M}\langle n-1|S|n \rangle_{M}|^2}{\sum_{i}(i+1)|{}_{M}\langle i|S|n \rangle_{M}|^2} \leq 1~.
\end{equation} 
It would be interesting to consider if one could modify the HM proposal 
so that the information loss does not occur for the Unruh excited state. 
The answer could be yes but since the Unruh vacuum state $|\Phi_0 \rangle_{in \otimes out}$  
and the excite state $|\Phi_1 \rangle_{in \otimes out}$  are not unitarily equivalent to each other, 
then the information loss may occur for the Unruh vacuum state under a new HM proposal, i.e.,
the final-state boundary condition.

Now, we would like to see whether the introduction of the Unruh excited state 
would cause any regularization problem. In order to do that we calculate
${}_{in \otimes out}\langle\Phi_0|b^{\dagger}_{out}b_{out}|\Phi_0\rangle_{in \otimes out}~$
and ${}_{in \otimes out}\langle\Phi_1|b^{\dagger}_{out}b_{out}|\Phi_1\rangle_{in \otimes out}~$, the latter is given by
\begin{eqnarray}
&   & {}_{in \otimes out}\langle\Phi_1|b^{\dagger}_{out}b_{out}|\Phi_1\rangle_{in \otimes out}  \nonumber \\
& = & \frac{2N}{N^2-1} {}_{in \otimes out}\langle\Phi_0|b_{out}b^{\dagger}_{out}b_{out}b^{\dagger}_{out}|\Phi_0\rangle_{in \otimes out} \nonumber \\
& = & \frac{2N}{N^2-1} \{ {}_{in \otimes out}\langle\Phi_0|\Phi_0\rangle_{in \otimes out}+2{}_{in \otimes out}\langle\Phi_0|b^{\dagger}_{out}b_{out}|\Phi_0\rangle_{in \otimes out} \nonumber \\
&   & +{}_{in \otimes out}\langle\Phi_0|(b^{\dagger}_{out}b_{out})^2|\Phi_0\rangle_{in \otimes out} \}.
\end{eqnarray}
By the way,

\begin{equation}
{}_{in \otimes out}\langle\Phi_0|b^{\dagger}_{out}b_{out}|\Phi_0\rangle_{in \otimes out}=\frac{1}{N} \sum_{i}i=\frac{N+1}{2},
\end{equation}
and
\begin{equation}
{}_{in \otimes out}\langle\Phi_0|(b^{\dagger}_{out}b_{out})^2|\Phi_0\rangle_{in \otimes out}=\frac{(2N+1)(N+1)}{6}.
\end{equation}
Substituting eqs. (25) and (26) into eq. (24), we obtain
\begin{equation}
{}_{in \otimes out}\langle\Phi_1|b^{\dagger}_{out}b_{out}|\Phi_1\rangle_{in \otimes out}=\frac{N(2N^2+9N+13)}{3(N^2-1)}\approx \frac{2N}{3}.
\end{equation}
Comparing eqs. (25) and (27), we can see that the Unruh excited state does not cause any additional regularization problem 
as compared with the Unruh vacuum state at least within the mathematical frame we employed in this paper.

Eqs. (22) and (23) indicate that the presence of matter near the event horizon would suppress the transfer of information 
from the collapsing matter to the outgoing Hawking radiation. As a result the evaporation of black hole may be affected by
the boundary condition as well. 

The Unruh vacuum state can be described by the condition that a free fall observer crossing the horizon long after the black hole forms would presumably see no very high positive free fall frequency excitations \cite{Jacobson}. On the other hand, the Unruh excited state turns out to be the final state of the system corresponding to the state starting initially with particles, very near the instant of the black hole formation, first studied by Wald \cite{Wald2}. The Unruh excited state defined by eq. (8) is mathematically equivalent to eq.(2.12) of the reference \cite{Wald2}.

From eqs. (7), (25) and (27), the Unruh excited state at the future null infinity in a micro-canonical form seems to contain a particle of finite energy. Thus, we can see that the HM proposal, especially the 
final state boundary condition, also allows the Unruh excited state as a boundary condition and is consistent with the
previous analysis \cite{Wald2, Muller}.

Unfortunately, changing the Hawking boundary condition of a black hole from the Unruh vacuum state to the Unruh excited state would be nontrivial and may require extremely high-energy excitations nonetheless. Certainly, this is beyond the capability of near future civilization. On the other hand, in quantum optics, the excitation of the Unruh excited state can be done by the single photon excitation of the two-mode squeezed state \cite{Barnett}. 
The primordial black holes in the early universe submerged in a dense soup of high energy particles might have the Unruh excited state as the HBC and as a result have longer lifetime than the lifetime predicted by Hawking \cite{hawk2}. Those surviving primordial black holes may be part of the dark matter in our universe. 
It has not escaped the author's notice that equations (4) and (9) also suggest the information exchange 
with a black hole \cite{Thorne} may not be strictly forbidden, in principle, by modulating the HBC.

\acknowledgements

This work was supported by KOSEF and MOST through the Creative Research Initiatives Program R-16-1998-009-01001-0(2006). The author thanks M. S. Kim for drawing his attention to this issue and valuable discussions

\end{document}